\documentclass[letterpaper, 10 pt, conference]{template/ieeeconf}  

\IEEEoverridecommandlockouts                              

\usepackage{graphics} 
\usepackage{amsmath} 
\usepackage{graphicx}
\usepackage{gensymb}

\title{\LARGE \bf
Multi-Agent Coordination Through Stigmergy for \\Canal Network Formation on Real Terrain
}

\author{
Jordan Kennedy$^{1}$, Federico Oliva$^{2}$, Amir Degani$^{2}$, Karen-Lee Bar Sinai$^{3}$, and Nare Karapetyan$^{1}$%
\thanks{$^{1}$Jordan Kennedy and Nare Karapetyan are with Woods Hole Oceanographic Institution (WHOI), Woods Hole, MA 02543, USA
{\tt\small \{jordan.kennedy, nare.karapetyan\}@whoi.edu}}%
\thanks{$^{2}$Federico Oliva and Amir Degani are with the Civil and Environmental Engineering Department, Technion -- Israel Institute of Technology, Haifa, 3200003, Israel
{\tt\small federico.oliva.it@gmail.com, adegani@technion.ac.il}}%
\thanks{$^{3}$Karen-Lee Bar Sinai is with the Faculty of Landscape Architecture, Graduate School of Design, Harvard University, Cambridge, MA 02138, USA
{\tt\small kl\_barsinai@gsd.harvard.edu}}%
}

\begin{document}
\maketitle

\thispagestyle{empty}
\pagestyle{empty}

\begin{abstract}

Terrain-modifying robot swarms operating in unstructured outdoor environments must coordinate under uncertain dynamics despite unreliable communication and incomplete global information. Stigmergy offers a decentralized alternative by allowing agents to coordinate indirectly through persistent environmental modifications, yet its application to large-scale terrain construction on real landscapes remains largely unexplored.
Inspired by beaver canal construction, we present a minimal stigmergic framework for multi-agent systems in which decentralized agents follow local wetness cues, excavate terrain, and interact through a closed agent--environmental feedback loop. The framework couples agents with seasonal hydrology on a real digital elevation model (DEM). 
Across 110 simulations on an $862$\,m$\times 1167$\,m region spanning agent count and hydrological recession rate, the model produces emergent canal-like networks whose morphology varies predictably with environmental constraints. An alignment-invariant morphological evaluation against a field-annotated reference network identifies parameter regimes that reproduce key canal characteristics, while failure cases reveal over-canalization and fragmentation. These results demonstrate that simple local terrain-modifying behaviors can generate plausible landscape-scale structures through environmental feedback, providing a minimal framework for studying decentralized terrain construction and bio-inspired swarm coordination.


\end{abstract}


\section{INTRODUCTION}

Multi-agent systems are increasingly deployed in outdoor environments~\cite{oliveira2021advances} with changing conditions where communication can be unreliable and global maps unavailable. In such conditions, centralized planning is often impractical, motivating communication-constrained decentralized approaches~\cite{karapetyan2017efficient}. One such strategy is \emph{stigmergy}~\cite{grasse1959reconstruction}, in which agents coordinate indirectly by modifying a shared environment~\cite{theraulaz1999brief}. This principle has inspired numerous swarm robotics systems, where a central challenge is explaining how coherent large-scale structure can emerge from local reactive interactions~\cite{werfel2014designing}.

\smallskip
Natural systems provide some insight into this process. Beavers (\textit{Castor canadensis}) build dams and canals that reshape the hydrology over entire landscapes without global planning or explicit coordination~\cite{naiman1988alteration,brazier2021beaver}. Rather than constructing a predefined network, individual beavers respond to local environmental conditions, while their persistent modifications alter future environmental conditions and consequently influence subsequent behavior. Over time, this agent--environment feedback produces extensive canal networks that remain long after the animals have left~\cite{grudzinski2020beaver}, making them a compelling example of stigmergic terrain construction.

\begin{figure}[t]
    \centering
    \includegraphics[width=\columnwidth]{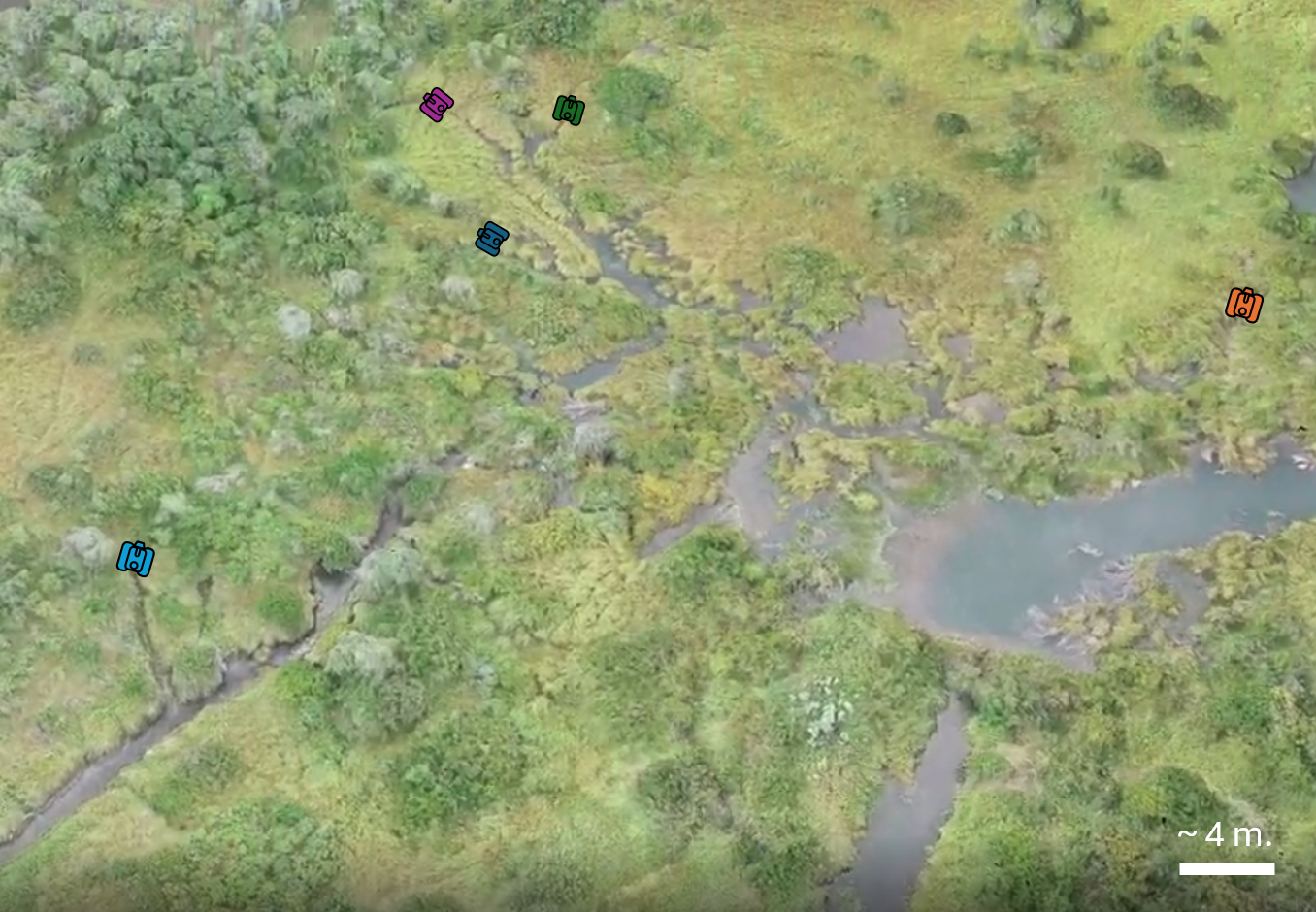}
    \caption{\textbf{Excavator Beaver Robots}. Conceptual deployment of agents (colored markers), operating under the proposed stigmergic framework.
    The background is reconstructed from drone imagery of an active beaver damming complex at the study site. 
    \vspace*{-15pt}}
    \label{fig:excavatorbots}
\end{figure}

\smallskip
Inspired by this process, we investigate whether a similarly minimal mechanism can generate canal-like landscape structures (Fig.~\ref{fig:excavatorbots}). We focus on one component of beaver landscape engineering: hydrology-driven canal-network formation. We do not model the full behavioral repertoire of beavers, nor do we claim that canal morphology is produced by their behavior alone. Instead, we ask a narrower generative question: \emph{Can decentralized terrain-modifying agents, using only local wetness cues and persistent excavation traces, produce canal-like networks on real terrain?} 

\begin{figure*}[t]
       \centering
    \includegraphics[width=1.0\textwidth]{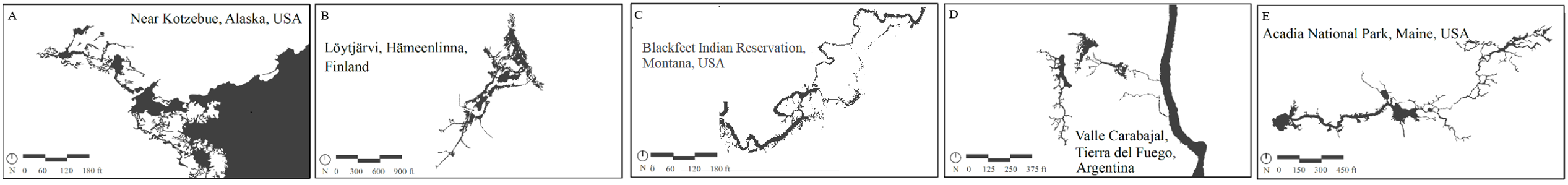}
    \caption{\textbf{Beaver damming complexes}. Human-annotated traces of canals of beaver damming complexes across ecosystems around the world and wetland types: (A) Near Kotzebue, Alaska, USA/Tundra permamelt; (B) Löytjärvi (Hämeenlinna), Finland/Forest Catchment lake; (C) Blackfeet Nation, Montana, USA./Great Plains headwater stream and representative simulation site; (D) Valle Carabajal, Tierra del Fuego, Argentina/ high elevation sub-Antarctic peatlands \& forests ;and (E) Acadia National Park, Maine, USA./Eastern deciduous forests first-order stream.  These examples illustrate the cross-landscape occurrence of canal networks; only site C is used for quantitative model evaluation in this study. Map credit: Gabriel Garon.} \vspace*{-18pt}
    \label{fig:dammingcomplexes} 
\end{figure*}


\smallskip
To study this question, we develop a closed-loop agent--environment framework in which the environmental cue is \emph{water}, represented by a dynamic wetness mask derived from drainage area and seasonal recession. The agent's action is \emph{excavation}, represented by cumulative terrain lowering that persists across simulation years. At the end of each simulated year, modified terrain is passed back into a hydrological model, closing an agent--environment feedback loop in which excavation can reinforce drainage pathways over multi-year timescales.

\smallskip
The main contributions of this work are:

\begin{enumerate}
    \item \textbf{A minimal stigmergic framework} for decentralized terrain modification, in which excavation and seasonal hydrology form a closed agent-environment loop on real terrain. 
    \item \textbf{A landscape-scale simulation pipeline} that couples decentralized excavation with annual hydrology recomputation on a one-meter DEM, enabling the study of canal formation.
    \item \textbf{An alignment-invariant morphological evaluation pipeline} that quantifies the similarity between simulated and field-annotated canal networks while evaluating the effects of agent count and hydrological recession.
\end{enumerate}

\smallskip

\vspace{-4mm}
\section{Related Work}

\subsection{Stigmergy and swarm robotics}

Stigmergy treats the environment as an active medium for coordination: agents modify their surroundings, and those modifications subsequently guide future behavior, thereby producing collective organization without centralized control \cite{grasse1959reconstruction, theraulaz1999brief}. Swarm robotics has instantiated this principle through physical traces such as blocks and surface markings, as well as virtual substrates such as pheromone fields and digital gradients, enabling distributed exploration, path formation, and construction \cite{petersen2019review, werfel2014designing}.

\smallskip
Prior stigmergic construction systems show that local rules can coordinate robot collectives through environmental traces. In many cases, however, the desired final structure is specified in advance: agents coordinate locally, but the target morphology is externally prescribed. The TERMES system \cite{werfel2014designing}, for example, demonstrated termite-inspired robots building user-specified 3D structures without explicit inter-agent communication, while extended stigmergy further showed how the environment can store information that improves robustness and parallelism \cite{werfel2006extended}. Related work in multi-robot excavation has addressed cooperative terrain modification toward target topographies, including planetary infrastructure preparation and decentralized earthmoving algorithms for user-specified terrain profiles \cite{thangavelautham2017lunar, ericland2025earthmover}. In contrast, the present model gives agents no desired canal map or target topography. Canal morphology generates from feedback: agents read hydrological cues, modify terrain locally, and then engage the new hydrological field produced by that process.

\subsection{Beaver behavior and canal formation}

Beavers are widely studied as ecosystem engineers, with extensive work on dam placement, impoundment dynamics, and landscape-scale geomorphic effects \cite{naiman1988alteration, brazier2021beaver}. Canal construction has received less attention, despite canals functioning as persistent hydrological infrastructure that concentrates flow, extends foraging range, and can remain visible across colony turnover \cite{grudzinski2020beaver} (Fig.~\ref{fig:dammingcomplexes}). This persistence makes canals a useful model for sematectonic stigmergy: the signal is a durable topographic modification that structures future movement.

\smallskip
Real canal networks reflect interacting processes including beaver activity, seasonal hydrology, sedimentation, vegetation, disturbance, bank erosion, and colony abandonment or reoccupation. The present model isolates one mechanism within that larger system: reinforcement of drainage pathways through repeated water-tracking excavation. This reduction tests whether canal-like morphology can arise from local wetness cues and persistent terrain modification alone.

Fieldwork \cite{kennedy2024cues} has shown that environmental cues shape the timing and location of construction activity, along with water conditions and existing canals, which together serve as major behavioral drivers. Here, those observations are formalized as a water-cue-driven stigmergic process grounded in real terrain data from the same study system. The field-annotated canal network is treated not as a causal record to reconstruct exactly, but as a morphological benchmark. Simulated and observed networks are compared by segment length, tortuosity, orientation, branch density, and fractal dimension. Thus, mismatch indicates which aspects of canal morphology may require additional mechanisms, e.g., vegetation-seeking behavior, sediment transport, storm disturbance, or dam--pond feedback.

%


%
 \begin{figure*}[t]
     \centering
     \includegraphics[width=0.9\textwidth,trim=0 5pt 0 0, clip]{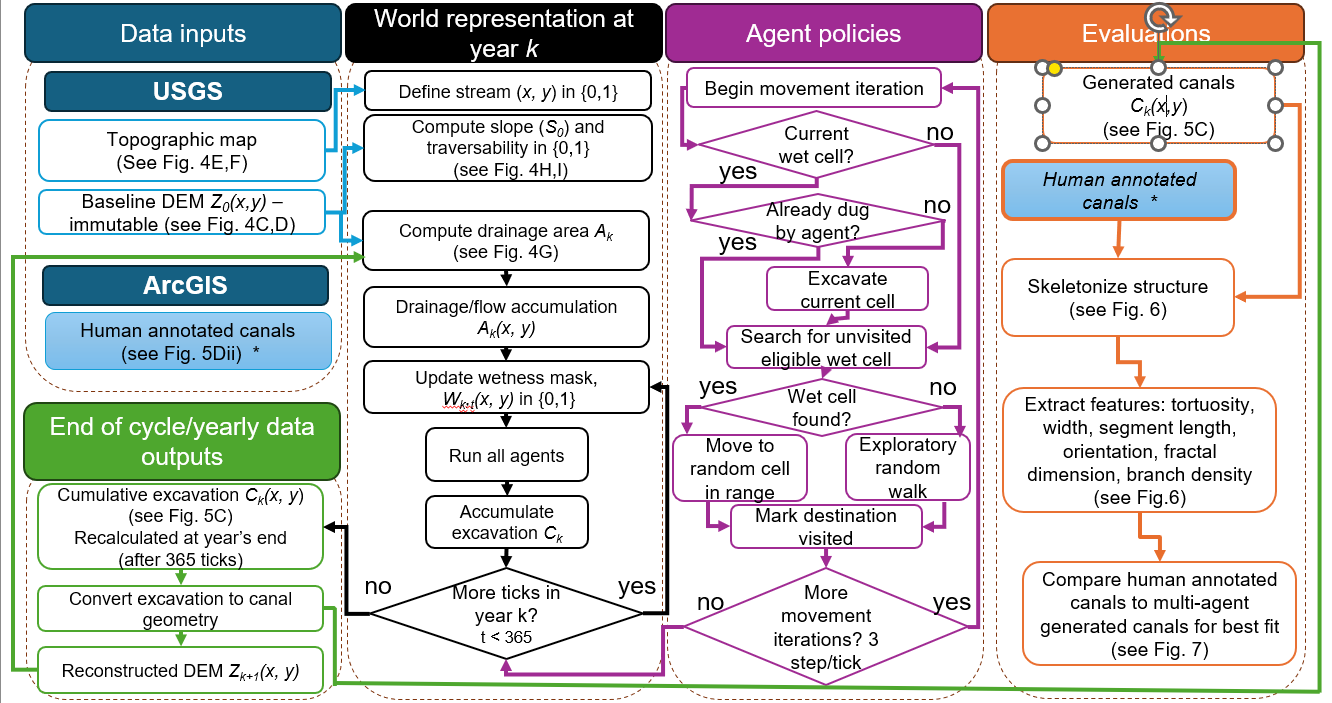}
     \caption{\textbf{Closed-loop agent--environment model.} Each simulated year $k$ consists of 365 ticks $t$. At every tick, the wetness field $W_{k,t}(x,y)\in\{0,1\}$ is updated using a drainage-modulated logarithmic drawdown, $W_{k,t}=f(t;r,A_k)$. In this seasonally retreating wet boundary in which cells with larger contributing areas $A_k(x,y)$ remain wet longer, agents begin each year on the stream and sense $\{W_{k,t},A_k,S_0\}$, where $S_0$ is calculated once from the baseline $Z_0$ and held fixed. A cell is traversable when $S_0(x,y)\leq\theta_{\max}$ and is a candidate destination when it is both traversable and wet. Agents move to the neighboring candidate cell; when no candidate is available, they explore  traversable cells until candidate is encountered. Each visit to a candidate cell adds an excavation increment $\delta$. Excavation accumulated during year $k$, denoted $\Delta C_k$, is added to the cumulative excavation field such that $C_{k+1}=C_k+\Delta C_k$. Years end $k$, $C_{k+1}$ is converted into saturated hemispherical excavation radii to reconstruct elevation surface $Z_{k+1}$ deterministically from the baseline $Z_0$. Hydrologic terrain analysis is then rerun on $Z_{k+1}$ to calculate updated drainage-area field $A_{k+1}$.\vspace*{-18pt}}
     \label{fig:logic}
 \end{figure*}

\section{Problem Statement}

Given a real landscape represented by gridded environmental layers, can decentralized excavation agents, using only local wetness cues and no global plan, generate persistent canal networks whose spatial structure resembles that of an observed beaver canal system?

\smallskip
The core hypothesis is behavioral. Beavers do not design canals. Instead, they respond to dynamic hydrological cues: seasonal flooding concentrates wetness along drainage corridors, agents track those wet cells, and repeated passage incises the terrain. That incision then concentrates subsequent flow, prolongs wetness along established corridors, and re-attracts agents in future seasons. Canal networks are therefore treated as the landscape-scale residue of agent-hydrology feedback, rather than as an intended design.

\smallskip
To test this hypothesis, the model must operate on real terrain rather than an abstract grid. We ground the simulation in a 1\,m resolution DEM of a study site in northwest Montana, using drainage accumulation as the primary hydrological proxy and seasonal recession to generate the retreating wet boundary that agents follow. Agents operate on strictly local information: they perceive nearby wetness, do not communicate, and excavate the cells they encounter as they track water. The rule set is intentionally minimal, to identify how local wetness tracking and persistent terrain modification alone can produce canal-like structure (Fig.~\ref{fig:excavatorbots}).

\smallskip
We developed this simulation as a reduced generative testbed for identifying where canal-like landscape modifications may emerge when beaver-inspired agents repeatedly follow and reinforce wet drainage corridors. The long-term goal is to support landscape management and restoration scenarios in which beaver reintroduction, coexistence planning, or bio-inspired terrain-modifying robots require forecasts of where water-seeking excavation is likely to reorganize a site.



\section{Proposed Framework}

The proposed framework consists of four components: (i) digital environment construction, (ii) environmental cue representation, (iii) decentralized agent behavior, and (iv) annual terrain and hydrological updates that close the agent–environment feedback loop (Fig.~\ref{fig:logic}).

\subsection{Digital environment construction}


The simulation is grounded in a digital representation of a real landscape with a spatial resolution of $1\,\mathrm{m}\times1\,\mathrm{m}$. Digital elevation models (DEMs) were obtained from the USGS National Map, while stream geometry was derived from USGS topographic data to define agent initialization locations. National Agriculture
Imagery Program (NAIP) imagery was used to derive vegetation quality indices, and drainage area was computed from the DEM to approximate persistent surface-water pathways.

\begin{figure*}[t]
    \centering
    \includegraphics[width=0.9\textwidth,trim=0 0pt 0 0, clip]{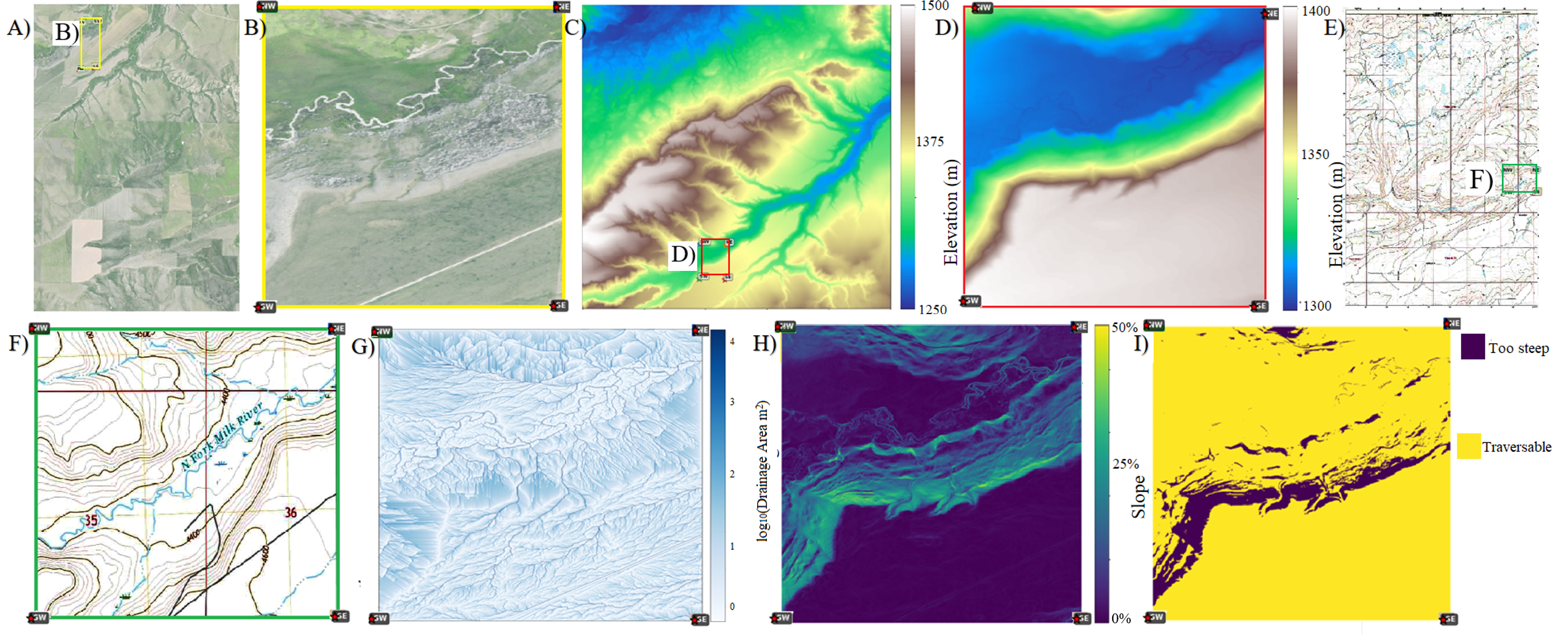}
    \caption{\textbf{Building a digital world}. (A) NAIP orthophoto imagery (GeoTIFF) was downloaded from the USGS and (B) cropped to a user-defined area of interest (latitude–longitude bounds). (C) Digital elevation models (DEMs) were downloaded from the USGS and (D) clipped to the same simulation domain. (E) A human-annotated USGS topographic map was used to delineate the stream network and (F) was clipped to the same simulation domain. (G) Drainage area (flow accumulation) was computed from the DEM to represent the drainage network. (H) Slope was derived from the DEM and (I) thresholded to generate a binary traversability mask for agent movement. Coordinates: NW = (-113.132491\degree,  48.932230\degree), SW = (-113.132491\degree,  48.921931\degree), SE = (-113.121137\degree,  48.921931\degree), NE = (-113.121137\degree,  48.932230\degree)\vspace*{-15pt}}
    \label{fig:wetmask_pipeline}
\end{figure*}

\smallskip
All raster datasets were reprojected to a common coordinate reference system and resampled to a common $1\,\mathrm{m}\times1\,\mathrm{m}$ raster
(Fig.~\ref{fig:wetmask_pipeline}). Because temporally matched datasets were not always available, layers were selected within a two-year acquisition window, which is appropriate given the decadal timescale of persistent beaver landscape
modification. The resulting raster layers form the simulation environment used by both the agents and the hydrological model.

\subsection{Environment representation and wetness cue model}

The simulation environment is represented as a collection of
co-registered raster layers defined over a common grid where each (x,y) cell is of $1\,\mathrm{m}\times1\,\mathrm{m}$ size. The simulation evolves over $Y$ annual cycles indexed by $k=1,\ldots,Y$. Each annual cycle consists of $T=365$ simulation ticks indexed by $t=1,\ldots,T$. The base topography is an immutable DEM defined as $Z_0$. The evolving landscape state consists of:

\begin{itemize}
    \item $A_k$: drainage area;
    \item $W_{k,t}$: binary wetness mask;
    \item $C_k$: cumulative excavation field;
    \item $Z_k$: terrain elevation (eroded DEM).
\end{itemize}

The terrain slope $S_0$, is calculated once and applied as a mask to limit agents to watershed for duration of simulation. The drainage-area raster $A_k$ and terrain elevation $Z_k$ are updated once per annual cycle, while the wetness mask $W_{k,t}$, and agent states evolve at every tick.  The wetness mask serves as the primary behavioral cue for the agents. Rather than explicitly modeling water depth,
discharge, groundwater exchange, sediment transport, or pond dynamics, drainage area is used as a first-order proxy for persistent surface-water availability. Cells with larger contributing areas remain wet longer during seasonal drawdown, whereas lower-drainage cells dry earlier. At each tick within a simulated year, the wetness mask is defined by thresholding the drainage area:
\begin{equation}
W_{k,t}(x,y)=
\begin{cases}
1, & A_{k}(x,y) \geq A_{\mathrm{crit}}(t),\\
0, & A_{k}(x,y) < A_{\mathrm{crit}}(t).
\end{cases}
\end{equation}

\smallskip
The seasonal threshold is
\begin{equation}
A_{\mathrm{crit}}(t)=
10^{
\log_{10}(A_{\min})+
r\frac{\log(t+1)}{\log(T+1)}
\left[\log_{10}(A_{\max})-\log_{10}(A_{\min})\right]
},
\end{equation}
where $A_{\min}$ and $A_{\max}$ are the minimum and maximum drainage-area values in the domain, and $r\in[0,1]$ is the recession-rate parameter. As $A_{\mathrm{crit}}(t)$ increases through the year, wetness contracts from broadly saturated lowlands toward high-drainage corridors. The recession rate $r$ is treated as an experimental regime parameter instead of a constant.

\subsection{Agent behavior}

Each agent follows a local, memoryless water-tracking policy consisting of search, navigation, and excavation (Fig.~\ref{fig:logic}). Agents are initialized on stream cells at the beginning of each annual cycle. At each simulation tick, an agent performs up to $v=3$ one-cell
micro-movements. For each micro-movement, an agent senses cells within the perception radius $p$ and constructs a candidate set
\begin{equation}
\mathcal{C} =
\{(x,y): W_{k,t}(x,y)=1,\; S_0(x,y)\leq \theta_{\max}\},
\end{equation}
where $\theta_{\max}=30^\circ$ is the maximum traversable slope. When one or more wet, traversable cells are detected within the perception radius,
the agent moves toward a candidate without reference to drainage area. If no wet
cells are detected, the agent performs a random walk restricted to traversable
cells until a wet cell is detected.

\smallskip
Targets are not retained across ticks. Instead, the candidate set is recomputed at every micro-movement, allowing agents to redirect as wetness recedes or as higher-drainage wet cells enter perception. If no candidate wet cell is detected, the agent takes a random exploratory step over traversable neighboring cells.

\smallskip
Excavation occurs when an agent occupies a wet, traversable cell. In the present implementation, excavation is represented as a visit-based activity increment rather than a time-resolved excavation process.
\begin{equation}
C_k(x,y) \leftarrow C_k(x,y)+\delta,
\end{equation}
where $\delta=0.05$\,m is the depth-equivalent excavation increment. Excavation does not impose an additional dwell-time cost; the counter is interpreted as cumulative activity intensity rather than literal instantaneous sediment removal. Agents have no memory, no communication, and no global perception. Coordination emerges only through shared access to the dynamic wetness mask and the excavation field.


\subsection{Annual cycle and hemisphere erosion operator}

At the end of each annual cycle, accumulated excavation is translated into canal geometry, a new eroded surface $Z_k$ is reconstructed, and hydrology is recomputed on that surface. This closes the feedback loop: excavation lowers terrain, lowered terrain concentrates drainage, concentrated drainage prolongs wetness, and persistent wetness re-attracts agents in later years. Canal cross-sections are approximated as hemispherical depressions. The cumulative excavation counter is converted to a canal radius using a saturation function:
\begin{equation}
R_k(x,y)=R_{\max}\left(1-e^{-\lambda C_k(x,y)}\right).
\end{equation}
Here, $R_{\max}$ limits the radius and depth of each erosion kernel, while $\lambda$ controls how quickly cumulative excavation approaches this limit. This scaling reduces excessive erosion, although superposed kernels may produce cumulative depths greater than $R_{\max}$. The $R_{\max}$ value was estimated from the measured maximum stream width at the study site. Lower values of $\lambda$ slow canal growth and were used conceptually to prevent unrealistically rapid deepening and widening. These parameters are not used by agents during movement; they are used only to reconstruct the carved DEM from the cumulative excavation field. For all results reported here, $R_{\max}=6  m$ and $\lambda \approx 1/R_{\max} = 1/6$. Given the radius field $R$ and immutable baseline DEM $Z_0$, the erosion depth contributed by a hemisphere centered at cell $(i,j)$ to a neighboring cell $(p,q)$ at Euclidean distance $d$ is
\begin{equation}
\Delta z_{(i,j)\rightarrow(p,q)} =
\begin{cases}
\sqrt{R_k(i,j)^2 - d^2}, & d \leq R_k(i,j),\\
0, & \text{otherwise.}
\end{cases}
\end{equation}

Contributions from all active cells are superposed, and the eroded DEM for year $k$ is
\begin{equation}
Z_k(x,y) = Z_0(x,y) -
\sum_{\{(i,j):R_k(i,j)>0\}}
\Delta z_{(i,j)\rightarrow(x,y)}.
\end{equation}
Because $Z_k$ is reconstructed from the immutable $Z_0$ and cumulative excavation field $C_k$, erosion is not repeatedly reapplied across annual cycles.

\subsection{Annual hydrology recomputation}

After erosion, drainage area is recomputed on $Z_k$ using D8 flow routing with depression handling, implemented with Landlab's \texttt{FlowAccumulator} and \texttt{DepressionFinderAndRouter} components~\cite{hobley2017landlab}. The updated field $A_{k+1}$ defines wetness conditions for the next annual cycle, while traversability remains defined by the fixed slope field $S_0$. Updated fields are written back to the simulation environment before the next year begins.

\begin{figure*}[t]
    \centering\includegraphics[width=0.85\textwidth,trim=0 0pt 0 0, clip]{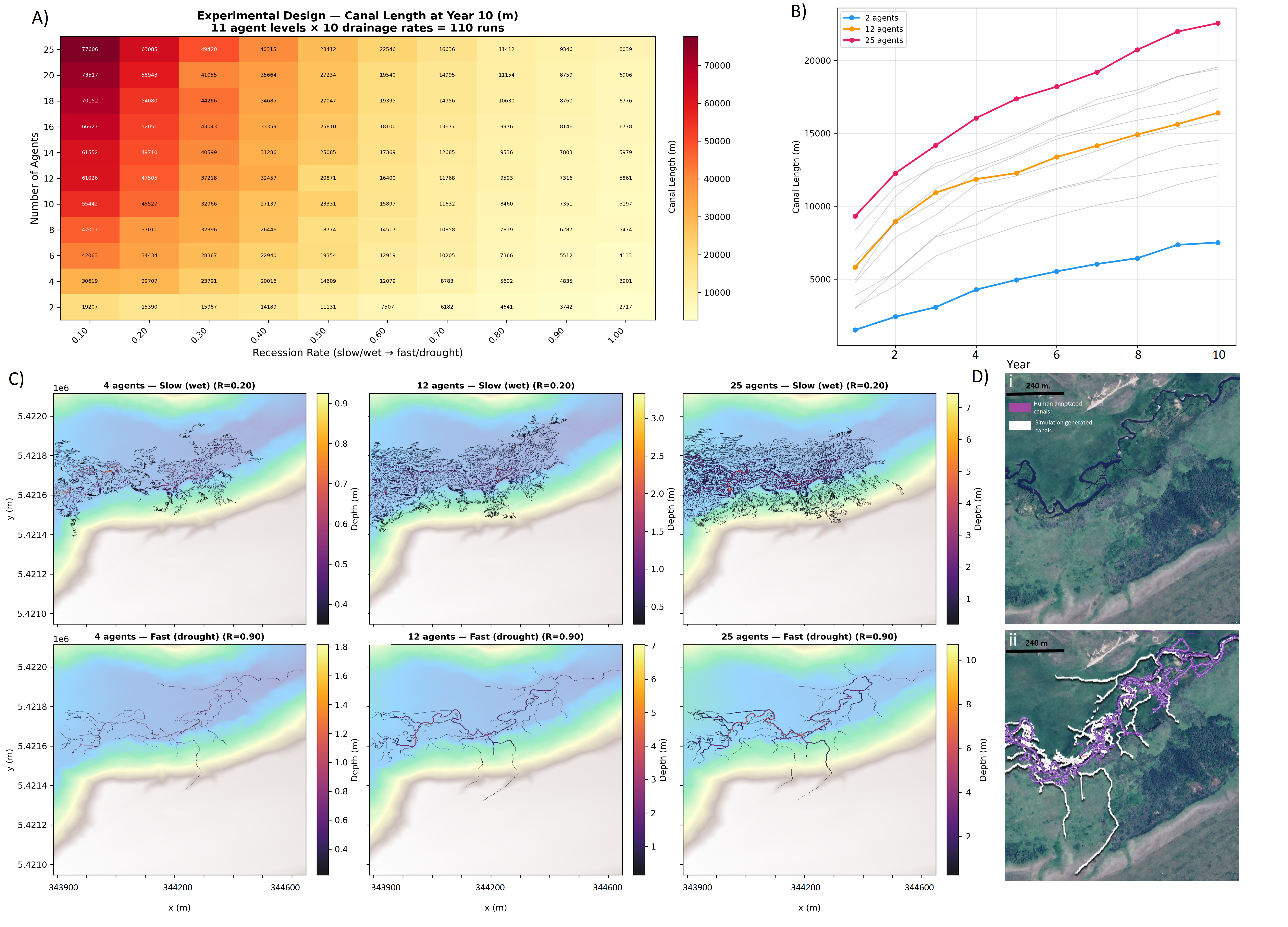}
    \caption{\textbf{Emergent canal morphology across hydrologic regimes and agent counts (10-year simulations).}
    \textbf{A)} Six parameter combinations: slow recession ($r=0.20$) with $N=4,12,25$, and fast recession ($r=0.90$) with $N=4,12,25$. All simulations use the same 1\,m DEM, 10 simulated years, 365 ticks/year, movement speed of 3 cells/tick, excavation increment $\delta=0.05$\,m, an expanding perception radius of 1--2 cells, traversability threshold $\theta_{\max}=30^\circ$, and fixed canal-radius conversion parameters.
    \textbf{B)} Example canal-length trajectories over 10 years at fixed $r$.
    \textbf{C)} Reconstructed canal-depth fields after 10 years (m; axes in UTM m), with the main stream in red and emergent canals in dark tones. Under slow recession ($r=0.20$), networks broaden and densify with increasing $N$; under fast recession ($r=0.90$), excavation concentrates into fewer, longer corridors, and increasing $N$ primarily deepens and extends the main branches. Color scales differ among panels.
    \textbf{D)} \textit{i}) Study-site imagery; \textit{ii}) simulated canals in white overlaid on human-annotated canals in purple.\vspace*{-15pt}}
\label{fig:emergence}
    
    \label{fig:emergence}
 \end{figure*}

\section{Implementation} 

The simulator is implemented in Python, using Mesa \cite{kazil2020mesa} for agent scheduling and behavioral execution, and Landlab \cite{hobley2017landlab} for annual hydrological recomputation. Within each annual cycle, Mesa agents read the current wetness mask, execute movement and excavation, and update the shared excavation field; Landlab then ingests the eroded DEM and recomputes drainage area for the following year. Traversability remains defined by the fixed baseline slope field $S_0$.

\smallskip
The simulator ingests the CSV world representation and executes $Y$ annual cycles. Each year, it writes three primary outputs: 

\begin{itemize}

\item \texttt{Cumulative excavation field} $C_k$, records depth-equivalent excavation activity per cell to date.

\item The derived \texttt{radius field} $R_k$, computed from $C_k$ via the saturation function and used to produce the eroded surface.

\item $Z_k$, \texttt{eroded DEM}, represents the modified landscape passed to Landlab for hydrological recomputation.

\end{itemize}

The separation between $C_k$ and $R_k$ is deliberate: retaining the cumulative excavation field decouples agent behavior from the canal geometry model, allowing the saturation parameters $R_{\max}$ and $\lambda$ to be adjusted in post-processing without re-running the full simulation. All outputs are georeferenced to the input CRS and grid.

\section{Experiments}

\subsection{Setup}

All simulations used the same base landscape, annual cycle, movement rule, and excavation rule. Each run lasted $Y=10$ annual cycles of $T=365$ ticks, representing a decade of cumulative activity. Fixed behavioral parameters were chosen to have movement speed $v=3$ cells per tick, perception radius $p\sim U(1,2)$ cells, maximum traversable slope $\theta_{\max}=30^\circ$ and visit-based excavation increment $\delta=0.05$\,m. Each run produced annual excavation, radius, and eroded DEM outputs.

\smallskip
The simulation contains three parameter classes. Behavioral parameters define how agents move and excavate: speed $v$, perception radius $p$, and excavation increment $\delta$. Environmental forcing is controlled by recession rate $r$, which determines how quickly wetness contracts from broadly saturated floodplain cells toward high-drainage corridors. Geometry-conversion parameters define how accumulated excavation is translated into canal form: $R_{\max}$ sets the maximum allowable canal radius, and $\lambda$ controls how rapidly repeated visits approach that radius.

\smallskip
We do not interpret $r$ as a measured drying rate, but rather as a dimensionless index of hydrological constraint strength. Low values represent persistent wetness across broad areas of the floodplain, while high values represent rapid contraction of wetness toward high-drainage corridors. Similarly, agent count $N$ is interpreted as excavation capacity rather than a direct census of individual animals: low $N$ represents limited repeated traffic, while high $N$, high excavation pressure. We conducted a full parameter sweep over agent count and recession rate:
\begin{itemize}
    \item Agent count $N \in \{2,4,6,8,10,12,14,16,18,20,25\}$
    \item Recession rate $r \in \{0.1,0.2,\ldots,1.0\}$
\end{itemize}
yielding $11 \times 10 = 110$ simulation runs (Fig. ~\ref{fig:emergence})

\section{Analyzing canal structure}

\begin{figure*}[t]
    \centering
    \includegraphics[width=0.9\textwidth]{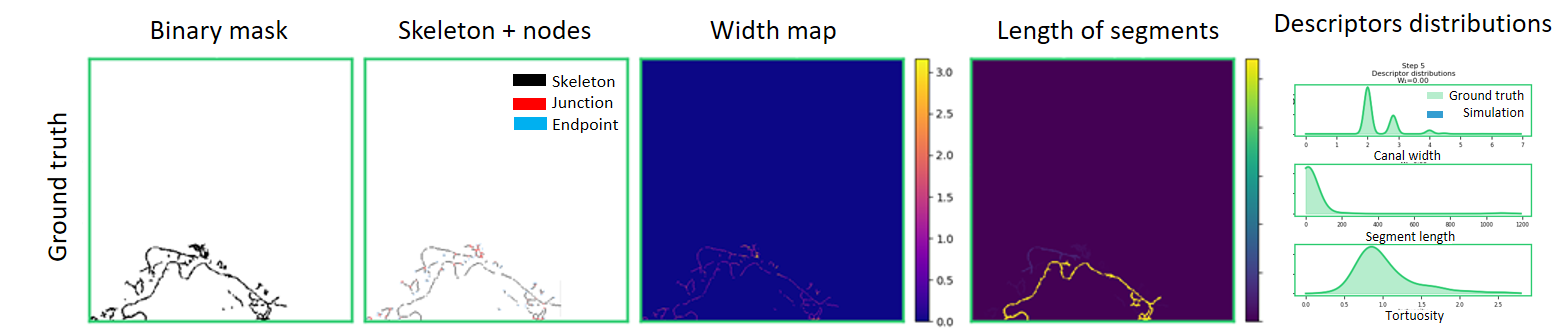}
    \caption{\textbf{Alignment-invariant morphological analysis.}
Pipeline: (1) binary canal mask; (2) centerline skeleton with junctions (red) and endpoints (blue); (3) width via distance transform sampled along the skeleton (radius, m); (4) connected components colored by segment length; (5) normalized width (solid) and length (dashed) distributions with Wasserstein-1 distances.\vspace*{-15pt}
}

    \label{fig:canalanlysis}
\end{figure*}

To compare simulated and annotated canal networks without requiring exact spatial overlap, we used an alignment-invariant morphological analysis (Fig.~\ref{fig:canalanlysis}). We reduced binary canal masks to one-pixel-wide centerlines using Zhang-Suen thinning \cite{zhang1984}, and removed small connected components as annotation noise. We identified endpoints (degree = 1) and junctions (degree $\geq 3$) on each skeleton, traced pixel paths between nodes to define canal segments, and extracted length, tortuosity (path length divided by straight-line distance), orientation, and width. Width was estimated by sampling a distance transform of the original canal mask along the skeleton.


We then computed network-level descriptors selected to capture structural properties relevant to beaver canal systems: total skeleton length, segment count, branch density, mean width, mean tortuosity, and box-counting fractal dimension \cite{mandelbrot1982}. These descriptors summarize extent, branching complexity, channel geometry, sinuosity, and scaling structure.
We quantified distributional similarity between simulated and annotated networks using Wasserstein-1 distance \cite{villani2009} for width, segment length, and tortuosity. We evaluated orientation differences using the Kolmogorov--Smirnov statistic \cite{massey1951}. This statistic only detects broad differences in directional alignment between simulated and annotated canal networks; we do not interpret it as a complete directional statistical model of canal routing. We compared scalar descriptors, including branch density and fractal dimension, using relative error. We normalized per-metric distances and combined them into a composite morphology score, with lower values indicating closer structural agreement, independent of spatial registration. This aggregate score serves as a compact diagnostic for ranking parameter regimes across the sweep, but it is not intended as a universal or uniquely optimal measure of canal-network similarity. Thus, we report each component metric separately. Model interpretation is based on the per-metric breakdown and the aggregate ranking.

\section{Results}

\subsection{Emergent canal formation across hydrological regimes}

Across 110 ten-year simulations, canal-like structures emerged from repeated tracking of the retreating wet boundary by agents. Excavation accumulated preferentially along persistent wet corridors and extended headward over time as agents reinforced the same low-lying pathways. The resulting morphology varied systematically with recession regime and agent count. Under \textit{slow recession} ($r \leq 0.3$), wetness persisted broadly across the valley bottom, producing a wide moving boundary with many locally equivalent wet cells. Excavation was diffuse: agents spread effort across marginally wet paths, producing braided, shallowly incised signatures rather than distinct canal corridors. Increasing the agent count in this regime amplified total excavation without improving canal morphology. Under \textit{fast recession} ($r \geq 0.7$), wetness collapsed rapidly onto high-drainage corridors, creating a narrow and strongly constrained wet boundary.

\noindent Canalization became more targeted: agents converged onto fewer routes, producing trunk corridors and branching patterns that more closely match observed canal networks. Agents' count controlled corridor continuity; too few produced fragmented networks; too many produced over-dense canalization outside the primary drainage pathways.

\subsection{Comparison to a field-annotated reference network}

Simulated canal outputs were compared against a field-annotated canal reference network derived from RGB imagery. Because the reference network represents the cumulative outcome of multiple historical processes, we do not interpret the comparison as an exact reconstruction of the observed site. Instead, we treat it as a benchmark for assessing whether the reduced water-tracking excavation rule reproduces canal forms within the same structural space as observed canals. Exact reconstruction would require modeling historical colony location, flood events, sediment infill, vegetation distribution, and dam--pond dynamics, which are outside the scope of the present model.

\smallskip
Instead, we use the annotation as a quantitative morphological benchmark. The purpose of the comparison is to test whether the reduced water-tracking excavation rule produces canal networks that fall within the same structural space as observed canals, rather than reproducing their exact spatial arrangement. For this reason, evaluation emphasizes morphology-based and alignment-invariant descriptors rather than exact spatial overlap.

\subsection{Geometric validation of simulated canal networks}

\begin{figure*}[t]
    \centering
    \includegraphics[width=0.9\textwidth]{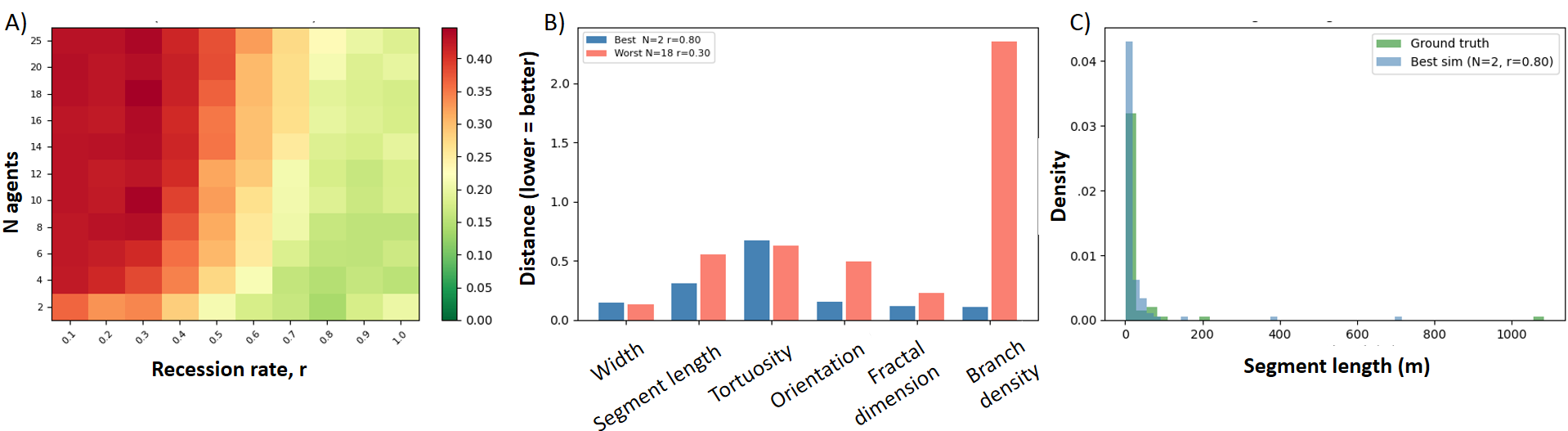}
    \caption{\textbf{Geometric shape comparison - Simulation vs Ground Truth (GT).} \textbf{A)} Aggregate geometric score (lower = more GT like).  \textbf{B)} Per-metric breakdown: best vs. worst. Color scales are standardized across panels to allow direct comparison of canal depth. \textbf{C)} Segment length distribution.\vspace*{-15pt}
}
    \label{fig:simanalysis}
\end{figure*}

Across the 110-run parameter sweep, the best-performing configuration was $N=2$, $r=0.80$, with an aggregate morphological score of 0.138, where lower values indicate closer agreement with the reference network (Fig.~\ref{fig:simanalysis}). This run reproduced the reference network's canal width and segment-length distributions most closely among the tested regimes. Its fractal-dimension error was 0.121, and its branch-density error was 0.109, indicating that the simulated network captured both scaling structure and branching intensity.

\smallskip
The primary residual mismatch was tortuosity, with a mean Wasserstein distance of 0.68. This suggests that wetness tracking alone does not fully reproduce the sinuosity of field-mapped canals, and that additional mechanisms may be needed to explain finer-scale path geometry. 

\smallskip
Model failures were also quantitatively identifiable. The worst-performing configuration, $N=18$, $r=0.30$, produced excessive branching, with a branch-density error approximately an order of magnitude larger than that of the best-performing runs. This failure mode corresponds to over-canalization: slow recession leaves wetness broadly distributed across the valley bottom, allowing agents to excavate many weakly constrained paths rather than converging on a sparse drainage-aligned network.

\smallskip
Together, the numerical comparison supports the visual pattern shown in Fig.~\ref{fig:simanalysis}: realistic canal morphology emerges only when hydrological recession strongly constrains the wet boundary and the agent count remains low enough to avoid excessive excavation.

\section{Discussion and Conclusions}

\subsection{Wetness constraint organizes excavation}

Within the model, the recession rate $r$ controls how quickly the wetness mask contracts from broad floodplain saturation toward high-drainage corridors. It should be interpreted as an index of the strength of environmental constraints, not as a calibrated field-drying rate. Low $r$ leaves many wet cells available, resulting in diffuse excavation and overbranched networks. High $r$ reduces the candidate set, forcing agents to converge on fewer drainage-aligned routes and producing clearer trunk corridors and headward extension. Thus, canal morphology improves when the behavioral cue field is spatially constrained enough to concentrate repeated excavation.

This result supports a model-specific mechanism: hydrological constraint can organize stigmergic excavation. It does not imply that recession alone determines real beaver canal morphology. In real landscapes, wetness interacts with vegetation, dam and pond configuration, sedimentation, flood disturbance, colony size, and prior canal traces. Across the tested parameter sweep, morphology scores were more sensitive to recession regime than to agent count, indicating that the structure of the wetness cue strongly conditions whether excavation becomes diffuse or canalized.

\subsection{Agent count controls excavation capacity}

Agent count serves as an excavation capacity parameter rather than a direct census of individual animals. Too few agents produce fragmented or weakly incised corridors, while moderate increases improve corridor continuity. Above a threshold, however, canal density becomes implausible: the network fills space, branches excessively, and extends outside core drainage pathways. This failure mode shows that maximizing excavation intensity or spatial coverage does not necessarily reproduce natural canal morphology.

\subsection{Limitations and implications}

In this single site study, the model is intentionally reduced. Agents respond only to local wetness, drainage area, and slope, while the wetness field serves as a behavioral cue rather than a full hydrodynamic simulation. Hydrology is recomputed annually, capturing seasonal landscape memory but not within-year flow dynamics, sediment transport, or calibrated water extent. Rather than reproducing the full complexity of beaver landscape engineering, it provides a minimal framework for studying how persistent environmental modifications can coordinate decentralized terrain-modifying agents. 


\smallskip
From the robotics perspective, this reframes terrain modification as an environmentally mediated construction problem, where large-scale structure emerges through interaction between agents and landscape rather than from a predefined blueprint.

\section*{Acknowledgments}
This work has been funded in part by the Woods Hole Oceanographic Institution (WHOI) through the Ladd Thorne Indigenous Knowledge Program, Carawan Fellowship and Investment in Science Program (ISP), and the Harvard University Center of Green Buildings, Cities and Graduate School of Design Faculty Research Funds. The authors thank Justin Werfel, William Boles, and Ben Witzel for their input.






\bibliographystyle{IEEEtran}
\bibliography{refs}

\end{document}